\documentclass[cameraready]{Interspeech}
\title{The First Environmental Sound Deepfake Detection Challenge:\\ Benchmarking Robustness, Evaluation, and Insights}

\author[affiliation={1}]{Han}{Yin}
\author[affiliation={2}]{Yang}{Xiao}
\author[affiliation={3}]{Rohan Kumar}{Das}
\author[affiliation={4,5}]{Jisheng}{Bai}
\author[affiliation={2}]{Ting}{Dang}

\address{
    $^1$ School of Electrical Engineering, KAIST, Daejeon, Republic of Korea \\
    $^2$ University of Melbourne, Australia, $^{3}$Fortemedia Singapore, Singapore,\\
    $^{4}$Xi’an University of Posts \& Telecommunications, Xi’an, China,\\
    $^{5}$Xi'an Lianfeng Acoustic Technologies Co., Ltd., China
}

\email{hanyin@kaist.ac.kr}


\usepackage{comment}
\usepackage{multirow}
\usepackage{booktabs}
\usepackage{makecell}
\usepackage[table]{xcolor} % 让表格支持颜色
\usepackage{xcolor}
\usepackage{makecell}
\usepackage{amssymb}
\usepackage{pifont}
\usepackage{multirow,balance}
\usepackage{subcaption}
\usepackage{array}
\usepackage{siunitx}
\usepackage{cite}
\hypersetup{hidelinks}
\usepackage{soul, xcolor}

\usepackage{soul, xcolor}

\begin{document}

\maketitle

\begin{abstract}
    Recent progress in audio generation has made it increasingly easy to create highly realistic environmental soundscapes, which can be misused to produce deceptive content, such as fake alarms, gunshots, and crowd sounds, raising concerns for public safety and trust. While deepfake detection for speech and singing voice has been extensively studied, environmental sound deepfake detection (ESDD) remains underexplored. To advance ESDD, the first edition of the ESDD challenge was launched, attracting 97 registered teams and receiving 1,748 valid submissions. This paper presents the task formulation, dataset construction, evaluation protocols, baseline systems, and key insights from the challenge results. Furthermore, we analyze common architectural choices and training strategies among top-performing systems. Finally, we discuss potential future research directions for ESDD, outlining key opportunities and open problems to guide subsequent studies in this field.
\end{abstract}

\section{Introduction}
Generative audio models for environmental sounds have advanced rapidly in recent years. State-of-the-art text-to-audio (TTA), audio-to-audio (ATA), and video-to-audio (VTA) systems can now synthesize rich and immersive soundscapes such as alarms, footsteps, sirens, and urban ambience~\cite{audioldm,audioldm2,diff-foley}. These capabilities enable various applications in media production, virtual reality, and accessibility. However, they also create a growing risk of malicious use, for example, in fabricating emergency scenarios, manipulating recordings, or amplifying misinformation via realistic but fake audio files~\cite{envsdd}. 

Compared with speech deepfake and anti-spoofing research, environmental sound deepfake detection (ESDD) faces unique challenges. Environmental sounds span a far more diverse set of sources and acoustic conditions, ranging from isolated events (e.g., a gunshot) to dense polyphonic soundscapes (e.g., a busy street or a train station). 
The presence of overlapping sound events and diverse spectral–temporal structures substantially increases the difficulty of designing robust detection systems. 
% \xy{Designing robust detection systems is difficult.}
In contrast to speech or singing voice deepfakes, where detectors often exploit pitch or phonetic inconsistencies, such cues may be absent or unreliable in environmental sounds due to the presence of diverse overlapping sound events. Therefore, detection strategies effective for speech or singing voice deepfakes may not generalize well to environmental sounds, highlighting the need for task-specific modeling approaches.

To address the lack of large-scale benchmarks for ESDD, EnvSDD~\cite{envsdd} was introduced as the first curated dataset for this task, containing 45.25 hours of real recordings and 316.7 hours of deepfake environmental sounds generated with multiple TTA and ATA models\footnote{\href{https://envsdd.github.io/}{The EnvSDD website: https://envsdd.github.io/}}. Experimental results on EnvSDD showed that, although in-domain detection is relatively easy, performance degrades substantially when test audio comes from unseen generators or unseen real-source datasets. Based on EnvSDD, the first ESDD challenge~\cite{esdd_eval} was developed to systematically benchmark robustness of ESDD systems under realistic conditions, comprising two distinct tracks\footnote{\href{https://sites.google.com/view/esdd-challenge/esdd-challenges/esdd-1/description}{The ESDD challenge page: https://sites.google.com/view/esdd-challenge/esdd-challenges/esdd-1/description}}~\cite{esdd_overview}. The introduction of the EnvSDD benchmark has rapidly drawn community attention to environmental audio anti-spoofing. 

% Although the organizers present a summary of the challenge results~\cite{esdd_overview}, the analysis remains preliminary. 

This paper presents a comprehensive overview of the first ESDD challenge, including the task formulation, dataset design, evaluation protocol, baseline systems, and a detailed analysis of top-ranking submissions. In addition to summarizing the leaderboard results, we examine system design trends, robustness characteristics, and key insights for future research\footnote{\href{https://github.com/apple-yinhan/ESDD-Review}{Code: https://github.com/apple-yinhan/ESDD-Review}}.

% including the evaluation protocol, the baseline systems, the detailed analysis of top-ranking systems, and open problems for future research\footnote{\href{https://anonymous.4open.science/r/ESDD_Review-14B4/README.md}{Anonymous repository link}}.

\section{Task, Database and Challenge}

\subsection{ESDD: Task at a Glance}

The goal of ESDD is to determine whether a given environmental sound clip originates from real-world recordings or has been generated or manipulated by artificial models. In contrast to speech anti-spoofing, which focuses primarily on human speech signals and a relatively narrow class of spoofing attacks, ESDD must handle a wide range of sound events and mixtures:
\begin{itemize}
    \item \textbf{Monophonic vs. polyphonic}: single-event clips (e.g., a dog bark) versus complex mixtures (e.g., street ambience with traffic, footsteps, and voices).
    \item \textbf{Generation paradigms}: TTA, ATA, and other modalities such as video-to-audio generation.
\end{itemize}

The challenge focuses on clip-level binary decisions (real vs. fake), with systems outputting continuous confidence scores for each 4-second segment.

\vspace{-1mm}
\subsection{EnvSDD Database}

EnvSDD is a large-scale dataset specifically designed for ESDD. Real audio clips are sampled from six public datasets covering both monophonic and polyphonic conditions, including UrbanSound8K~\cite{urbansound}, DCASE 2023 Task 7 Dev~\cite{dcase2023task7}, TAU 2019 Open Dev~\cite{tau_uas}, TUT SED 2016~\cite{mesaros2016tut}, TUT SED 2017~\cite{tut-2017-dev} and Clotho~\cite{drossos2020clotho}. All audio signals are resampled to 16 kHz and segmented into 4-second clips.

Deepfake clips are generated using five TTA models and two ATA models. The TTA models include AudioLDM~\cite{audioldm}, AudioLDM~2~\cite{audioldm2}, AudioGen~\cite{kreuk2022audiogen}, TangoFlux~\cite{hung2024tangoflux}, and AudioLCM~\cite{liu2024audiolcm}, while ATA generation is performed by AudioLDM and AudioLDM~2 conditioned on acoustic features. For monophonic audio, metadata labels are rewritten into simple captions, whereas for polyphonic audio, a large language model is used to synthesize descriptive captions from scene and event labels. These captions are then used as prompts for TTA models; while ATA models operate directly on acoustic representations.

\begin{figure}[t]
    \centering
    \begin{subfigure}[t]{\linewidth}
        \centering
        \includegraphics[width=0.9\linewidth]{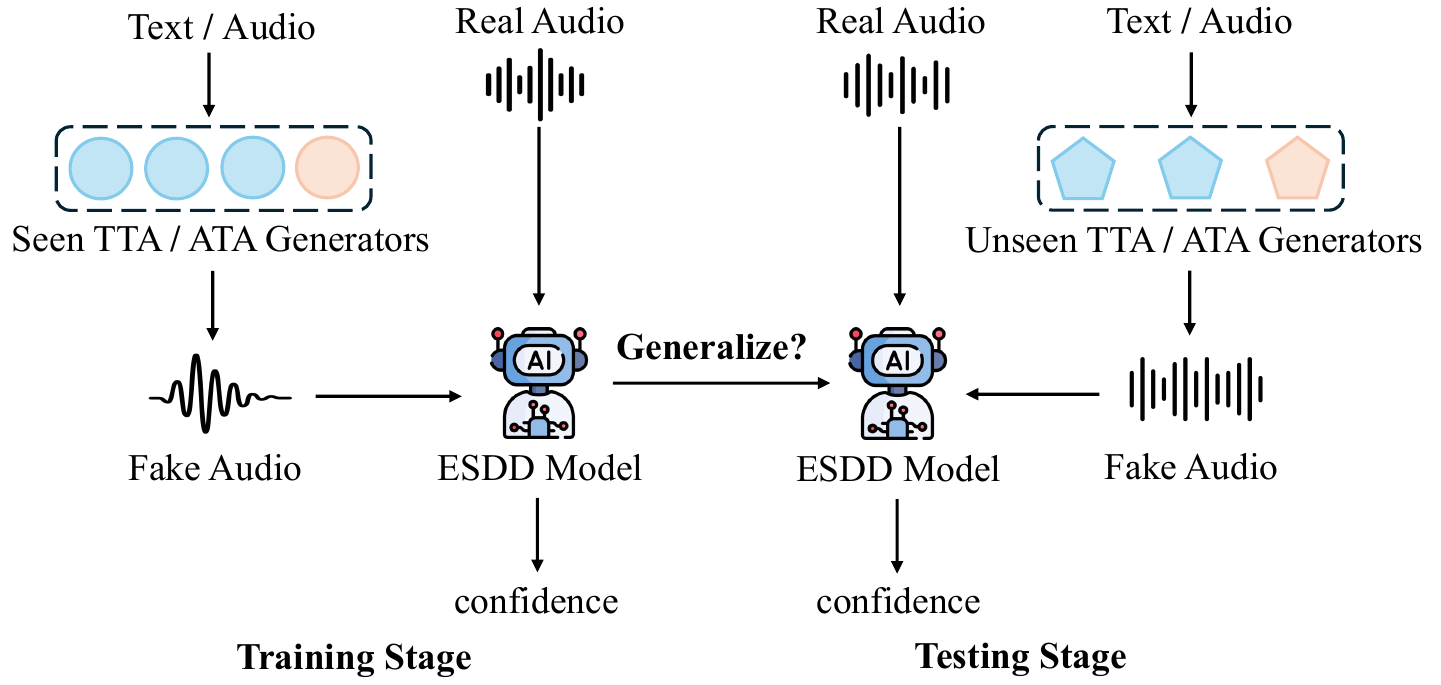}
        \caption{Track 1: ESDD in Unseen Generators.}
        \label{fig:track1}
    \end{subfigure}

    \vspace{2mm} % adjust (or use a negative value to tighten)
    \begin{subfigure}[t]{\linewidth}
        \centering
        \includegraphics[width=0.9\linewidth]{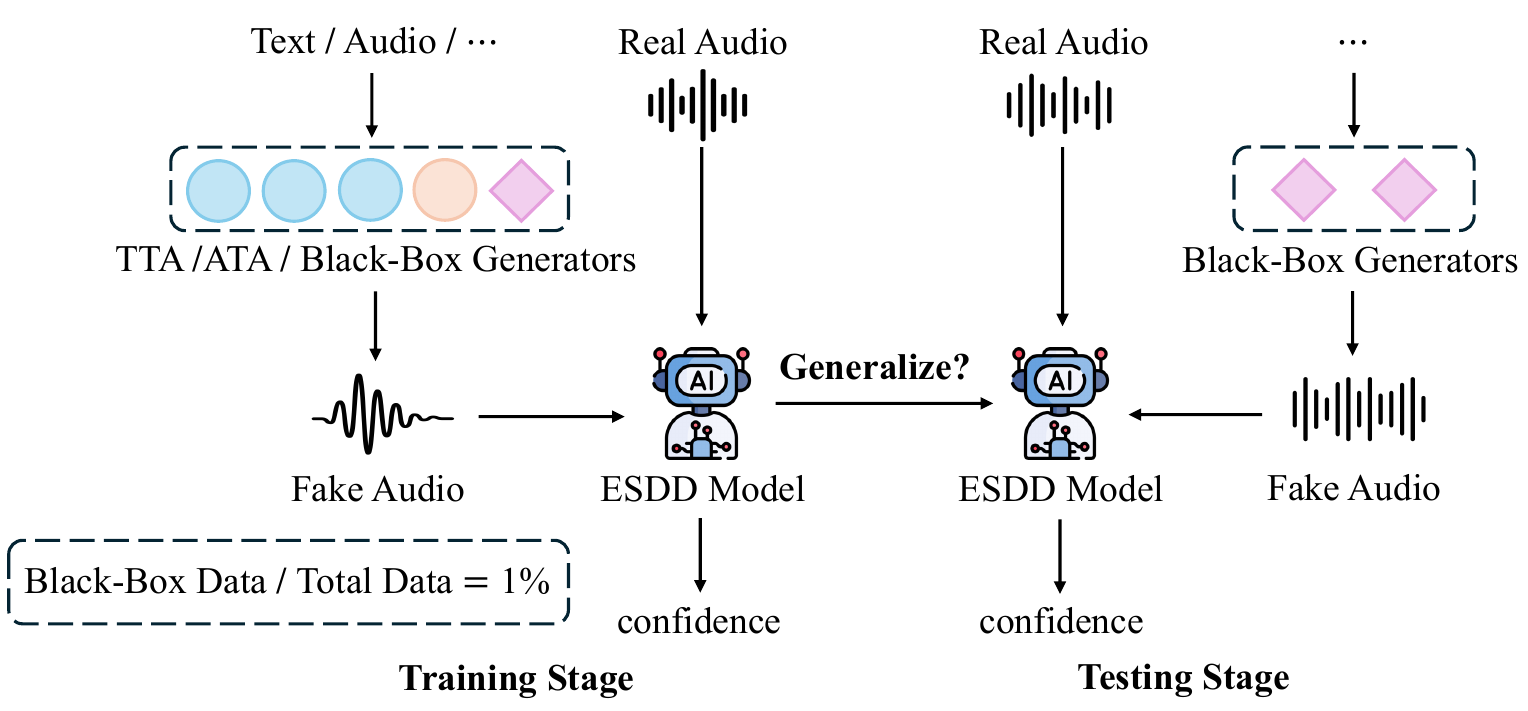}
        \caption{Track 2: Black-Box Low-Resource ESDD.}
        \label{fig:track2}
    \end{subfigure}
\vspace{-2mm}
    \caption{Overview of the two tracks fo ESDD challenge.}
    \label{fig:tracks}
    \vspace{-2em}
\end{figure}

\subsection{ESDD Challenge}
As shown in Figure~\ref{fig:tracks}, the challenge has two tracks: ESDD in Unseen Generators (Track 1) and Black-Box Low-Resource ESDD (Track 2), aiming to promote the development of detection methods that are both generalizable and data-efficient.

\subsubsection{Track 1: ESDD in Unseen Generators}
For Track~1, the challenge uses EnvSDD as the database, as shown in Table~\ref{tab:track_1}, training and validation data contain deepfake sound clips generated by a subset of models, while the test set uses different, unseen TTA and ATA generators. This setup is designed to explicitly assess cross-generator generalization, encouraging systems to learn generator-agnostic detection cues rather than overfitting to artifacts specific to particular synthesis models. Therefore, robust performance under this setting indicates a strong capability to capture inconsistencies between real and synthetic sound signals, improving the practical reliability of ESDD systems in open-world conditions.

\subsubsection{Track 2: Black-Box Low-Resource Data}

In Track 1, although the generators used in the test set are unseen, the underlying audio generation paradigms (i.e., TTA and ATA) are present during training. 
In Track 2, the challenge introduces a more challenging evaluation scenario in which the generation paradigm is neither TTA nor ATA. Participants are not provided with explicit information about the generation method used in the test data; it may involve any approach beyond TTA and ATA. This evaluation setting is termed the \textbf{black-box test}, as it is close to real-world deployment conditions where detection systems must operate reliably without prior knowledge of the underlying generation mechanisms. In addition, black-box data is often scarce and difficult to collect in real-world scenarios. Therefore, a low-resource setting is simulated by providing only a limited portion of such black-box data for training, corresponding to 1\% of the total development set under Track 2. This setting of \textbf{Black-Box Low-Resource ESDD} aims to assess the ability of detection models to generalize under severe data constraints while confronting entirely unknown generation paradigms.
Table~\ref{tab:track_2} summarizes the statistics of the black-box data. Compared to Track 1, the scale of the black-box dataset is substantially smaller. For Track 2, the challenge combines the training and validation sets from Track 1 with the black-box data, with the latter comprising only 1\% of the resulting development set.

As the challenge has now been completed, detailed construction of the black-box data is disclosed. Specifically, the deepfake sound clips in the black-box set are generated using VTA models. This design reflects realistic misuse scenarios in which sound is synthesized to match manipulated or fabricated video content, such as forged surveillance footage or staged online media. Two VTA models, namely FoleyCrafter~\cite{foleycrafter} and DiffFoley~\cite{diff-foley}, are used to generate fake sounds. For real video samples, clips are randomly selected from the VGGSound~\cite{chen2020vggsound} dataset, which contains diverse real-world sound events paired with video content, thereby ensuring broad coverage of environmental sound categories and acoustic conditions. 

\begin{table}[t]
\centering
\caption{Statistics of the dataset in Track 1.}
\vspace{-0.6mm}
\renewcommand\arraystretch{1}{
\setlength{\tabcolsep}{1.2mm}{
\scalebox{0.95}{
\begin{tabular}{c|ccccc}
\toprule
\multirow{2}{*}{\textbf{Usage}} & \multicolumn{2}{c}{\textbf{Real}} & \multicolumn{3}{c}{\textbf{Fake}} \\
 & Sources & \# Clips & TTA & ATA & \# Clips \\
\midrule
Training & \cite{urbansound,tau_uas,mesaros2016tut,tut-2017-dev} & 27,811 & \cite{audioldm,audioldm2,kreuk2022audiogen} & \cite{audioldm} & 111,244\\
Validation & \cite{urbansound,tau_uas,mesaros2016tut,tut-2017-dev} & 7,942 & \cite{audioldm,audioldm2,kreuk2022audiogen} & \cite{audioldm} & 31,768\\
Test & \cite{dcase2023task7,drossos2020clotho} & 1,000 & \cite{hung2024tangoflux, liu2024audiolcm} & \cite{audioldm2} & 3,000\\
\bottomrule
\end{tabular}}
}
}
\label{tab:track_1}
\vspace{-0.5em}
\end{table}

\begin{table}[t]
\centering
\caption{Statistics of the black-box data in Track 2.}
\vspace{-0.6mm}
\renewcommand\arraystretch{1}{
\setlength{\tabcolsep}{5.5mm}{
\scalebox{1}{
\begin{tabular}{ccc}
\toprule
\textbf{Usage} & \textbf{\# Real Clips} & \textbf{\# Fake Clips}\\
\midrule
Training & 270 & 1,083 \\
Validation & 90 & 361 \\
Test & 1,994 & 7,980 \\
\bottomrule
\end{tabular}}
}
}
\label{tab:track_2}
\vspace{-2em}
\end{table}

\subsubsection{Evaluation Metric and Challenge Baselines}

\noindent \textbf{Evaluation Metric:} The challenge adopts the Equal Error Rate (EER) as the evaluation metric for ESDD. A confidence score is estimated for each sound clip, representing the likelihood that the clip is real. By varying the decision threshold applied to these scores, a trade-off is obtained between the false acceptance rate (FAR) and the false rejection rate (FRR). The EER is defined as the operating point at which FAR and FRR are equal. A lower EER corresponds to better detection performance.

\noindent \textbf{Baselines:} The challenge considers two baseline systems: AASIST and BEATs+AASIST. AASIST~\cite{jung2022aasist} is an end-to-end architecture that employs a heterogeneous stacking graph attention mechanism to model acoustic representations.
Building upon this framework, BEATs+AASIST integrates BEATs~\cite{chen2023beats} as a front-end feature extractor to provide high-level acoustic representations. By leveraging pretrained audio representations, this hybrid system achieves improved performance over the vanilla AASIST model on the EnvSDD dataset.

\begin{table}[t]
\centering
\caption{Performance in EER (\%) of different detection systems in ESDD challenge Track 1. ``Ensem'' represents ``Ensemble''.}
\vspace{-2mm}
\renewcommand\arraystretch{0.8}{
\setlength{\tabcolsep}{0.5mm}{
\scalebox{1}{
\begin{tabular}{c|c|c|c|c}
\toprule
\textbf{Team} & \textbf{System} & \textbf{ID} & \textbf{Ensem} & \textbf{EER} \\
\midrule
AHU & EAT+AASIST & S01 & 5 & \textbf{0.30}\\
DFKI & EAT-L+BiCrossMamba & S02 & 4 & 0.80 \\
CUC & SSLAM+FFN & S03 &1 & 1.20 \\
CAU~\cite{cau} & BEAT2AASIST & S04 & 2 & 1.60 \\
CAU~\cite{cau} & BEAT2AASIST & S05 & 2 & 1.62 \\
CAU~\cite{cau} & BEAT2AASIST & S06 & 1 & 1.70 \\
BIT & EAT+ArcFace & S07 & 1 & 2.40 \\
BIT & DA-EAT-LA & S08 & 1 & 3.30 \\
BUT~\cite{but} & EAT-L+MHFA & S09 & 3 & 4.38 \\
BUT~\cite{but} & EAT+MHFA-DSU & S10 & 1 & 4.80 \\
BUT~\cite{but} & EAT+MHFA & S11 & 1 & 4.80 \\
HEU & BEATs+KNN & S12 & 1 & 4.90\\
IIT JAMMU & Phoenix's ESDD & S13 & 1 & 7.00\\
DKU & SSLs+AASIST & S14 & 3 & 7.30 \\
USTUTT & EAT+RawBoost+AASIST & S15 & 1 & 8.92 \\
USTUTT & EAT+AASIST & S16 & 1 & 10.80 \\
JAIST~\cite{jaist} & ATCA & S17 & 1 & 11.28 \\
 \rowcolor{blue!10}
Baseline 1~\cite{esdd_eval} & BEATs+AASIST & B01 & 1 & 13.20 \\
 \rowcolor{blue!10}
Baseline 2~\cite{esdd_eval} & AASIST & B02 & 1 & 15.02 \\
\bottomrule
\end{tabular}}
}
}
\label{tab:result_track_1}
\end{table}

\begin{table}[t!]
\centering
\caption{Performance in EER (\%) of different detection systems in ESDD challenge Track 2. ``Ensem'' represents ``Ensemble''.}
\vspace{-0.6mm}
\renewcommand\arraystretch{0.8}{
\setlength{\tabcolsep}{1.6mm}{
\scalebox{1}{
\begin{tabular}{c|c|c|c|c}
\toprule
\textbf{Team} & \textbf{System} & \textbf{ID} & \textbf{Ensem} & \textbf{EER} \\
\midrule
DFKI & \makecell{EAT-L-SSLAM+\\BiCrossMamba} & M01 & 5 & \textbf{0.25} \\
AHU & EAT+AASIST & M02 & 5 & \textbf{0.25} \\
CAU~\cite{cau} & BEAT2AASIST & M03 &3 & 0.35 \\
CAU~\cite{cau} & BEAT2AASIST & M04 &2 & 0.40 \\
CAU~\cite{cau} & BEAT2AASIST & M05 &3 & 0.41 \\
CUC & SSLAM+FFN & M06 &1 & 1.05 \\
HEU & BEATs+KNN & M07 &1 & 2.96 \\
 \rowcolor{blue!10}
Baseline 1~\cite{esdd_eval} & BEATs+AASIST & B01 &1 & 12.48 \\
JAIST~\cite{jaist} & ATCA & M08 &1 & 12.64 \\
 \rowcolor{blue!10}
Baseline 2~\cite{esdd_eval} & AASIST & B02 &1 & 15.40 \\
\bottomrule
\end{tabular}}
}
}
\label{tab:result_track_2}
\vspace{-1em}
\end{table}

\section{Challenge Results and Findings}
The challenge attracted substantial interest from both academia and industry. In total, 97 teams registered, and more than 1{,}700 valid submissions were received across the two tracks. A large majority of teams were affiliated with academic institutions, with additional participation from industrial research labs and mixed academic–industrial collaborations. Many teams reported prior experience with audio deepfake detection, especially in speech and singing domains, highlighting growing attention to environmental sound deepfakes.

\subsection{Overall Results and Analyses}

Table~\ref{tab:result_track_1} presents the results for Track 1. Among the baseline systems, BEATs+AASIST slightly outperforms AASIST, highlighting the benefit of incorporating high-level acoustic representations extracted from the self-supervised pre-trained model BEATs. The first ranked submission achieves an EER of 0.3\% on the test set, yielding an absolute improvement of 12.9\% compared to the BEATs+AASIST baseline.
Similar trends are observed in Track 2. As shown in Table~\ref{tab:result_track_2}, where the first ranked system presents an improvement of 12.23\% compared to BEATs+AASIST.
Here, we summarize \textbf{five main strategies} for system design proposed by the participants of this challenge.

\textbf{Acoustic Features from Pre-trained Models:} For Baseline 2, the challenge uses BEATs to extract acoustic features, which is beneficial for improving the model's generalization ability. CAU and HEU teams continue to use this method. The CAU team propose BEAT2AASIST, an enhanced architecture that splits BEATs-derived representations along frequency and channel dimensions and processes them through double AASIST branches. Other teams (e.g., AHU, DFKI, BIT, BUT) choose to use more recent pre-trained models, such as EAT \cite{eat} and SSLAM \cite{sslam}, which further improved the detection performance. DFKI introduces an attentive layer fusion mechanism that bridges shallow and deep pre-trained SSL features.

\textbf{Advanced Classification Model:} Some teams focus on designing modules that can help the classification back-end capture acoustic features more effectively. Specifically, DFKI employs BiCrossMamba, a bidirectional selective state space model tailored for spectro-temporal modeling. CAU explores multiple top-k transformer layer fusion strategies to enrich feature representation. JAIST applies a text-guided cross-attention model for classification, while BUT incorporates a lightweight Multi-Head Factorized Attention (MHFA) module to better capture discriminative latent representations.

\textbf{Data Augmentation:} To increase the model's robustness across different generators, data augmentation is applied by various teams.
AHU team introduces a targeted data augmentation protocol, incorporating semantically-aligned data construction, MP3 compression, and loudness normalization.
DFKI team utilizes a ``Cut and Mix'' data augmentation strategy and additional real training data from AudioCaps~\cite{audiocaps}.
CAU team augments the training set with vocoder-generated fake audio, while BUT introduces a feature domain augmentation strategy to enhance model robustness against unseen spectral distortions.

\textbf{Training Strategy:} In the challenge, the provided development set exhibits an unbalanced distribution, with substantially more fake samples than real ones. To address this issue, CUC applies a class-weighted training objective to handle the severe data imbalance and improve robustness to unseen TTA and ATA generators. In addition, BIT proposes a solution that combines LoRA~\cite{hu2022lora} fine-tuning, domain adversarial training~\cite{ganin2016domain} and ArcFace~\cite{deng2019arcface} loss for better generalization performance.

\textbf{Ensemble:} Several teams employ an ensemble approach for an improved detection performance. We explicitly report the number of individual systems involved in each ensemble in Table \ref{tab:result_track_1} and Table \ref{tab:result_track_2}, providing a detailed breakdown of the ensemble configurations used by different submissions. Overall, ensemble-based systems consistently outperform single-model approaches, achieving substantially lower EERs. 

\begin{table}[t]
\centering
\caption{Audio generators used in the ESDD challenge.}
\vspace{-0.6mm}
\renewcommand\arraystretch{0.8}{
\setlength{\tabcolsep}{1mm}{
\scalebox{1}{
\begin{tabular}{ccccccc}
\toprule
\multirow{2}{*}{\textbf{Generator}} & \multirow{2}{*}{\textbf{Method}} & \multirow{2}{*}{\textbf{ID}} & \multicolumn{2}{c}{\textbf{Track 1}} & \multicolumn{2}{c}{\textbf{Track 2}}\\
& & & Training & Test & Training & Test \\
\midrule
AudioLDM & TTA & G01 & \ding{52} & \ding{55} & \ding{52} & \ding{55}\\
AudioLDM2 & TTA & G02 & \ding{52} & \ding{55} & \ding{52} & \ding{55}\\
AudioGen & TTA & G03 & \ding{52} & \ding{55} & \ding{52} & \ding{55}\\
AudioLDM & ATA & G04 & \ding{52} & \ding{55} & \ding{52} & \ding{55}\\
AudioLCM & TTA & G05 & \ding{55} & \ding{52} & \ding{55} & \ding{55}\\
TangoFlux & TTA & G06 & \ding{55} & \ding{52} & \ding{55} & \ding{55}\\
AudioLDM2 & ATA & G07 & \ding{55} & \ding{52} & \ding{55} & \ding{55}\\
FoleyCrafter & VTA & G08 & \ding{55} & \ding{55} & \ding{55} & \ding{52} \\
DiffFoley & VTA & G09 & \ding{55} & \ding{55} & \ding{52} & \ding{52} \\
\bottomrule
\end{tabular}}
}
}
\label{tab:generators}
\vspace{-2em}
\end{table}

\begin{figure*}[t]
\centering
\includegraphics[width=2.1\columnwidth]{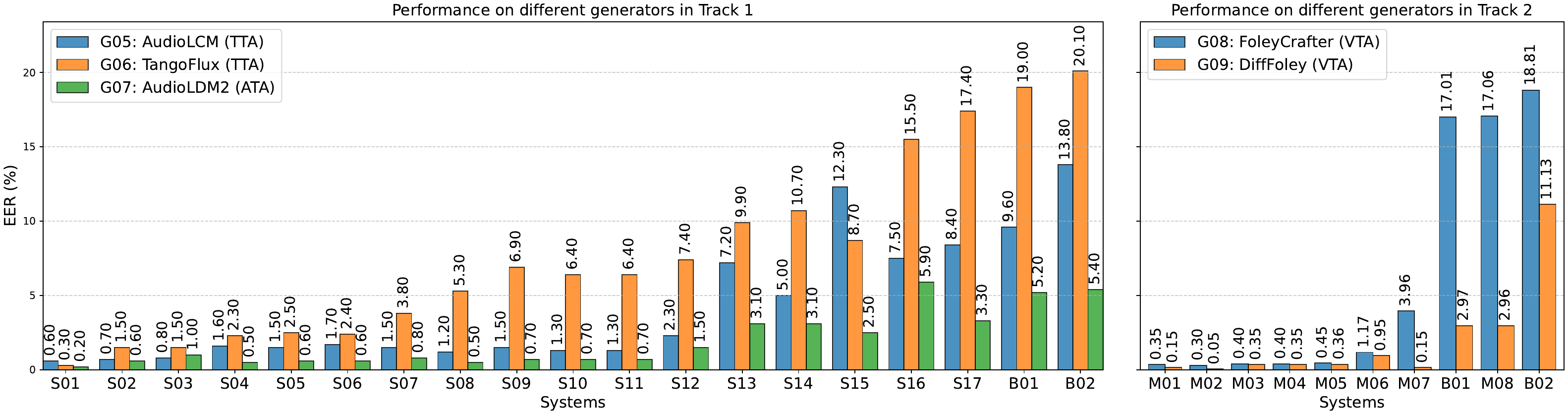}
\vspace{-2mm}
\caption{Deepfake detection performance in EER (\%) of different systems across various audio generators in Track 1 and Track 2.}
\label{fig:eers}
\vspace{-6mm}
\end{figure*}

\subsection{Performance on Different Generators}
In Table~\ref{tab:generators}, we provide a detailed table of all audio generation models involved in the ESDD Challenge.
Figure~\ref{fig:eers} illustrates the performance of all submitted systems across the different unseen generators for both Track 1 and Track 2. For Track 1, the systems are evaluated on deepfakes generated by AudioLCM (G05), TangoFlux (G06), and AudioLDM2 (G07). It can be observed that there is a substantial variation in EER across different generation paradigms.
Specifically, G06 (TangoFlux, a TTA model) emerges as the most challenging generator to detect. The baseline systems, B01 (BEATs+AASIST) and B02 (AASIST), suffer severe degradation when exposed to G06, yielding EERs of 19.00\% and 20.10\%, respectively. This indicates that the advanced flow-matching and preference optimization mechanisms employed by TangoFlux successfully minimize the synthetic artifacts typically captured by baseline front-ends. In contrast, top-performing systems demonstrate remarkable robustness against this challenging attack. For instance, the primary system from the AHU team (S01) achieves an exceptionally low EER of 0.30\% on G06. This suggests that the combination of semantically-aligned data augmentation and the ensemble of diverse classifiers effectively captures generator-agnostic spoofing cues that single-model baselines miss.

Comparing single-model performance with ensemble approaches further highlights the necessity of fusion strategies. The best single system (S03) achieves an EER of 1.50\% on G06, which is still five times higher than the ensemble S01. This discrepancy demonstrates that reliable generalization to high-fidelity, unseen TTA generators heavily depends on the complementary representations extracted by multiple sub-systems.

Conversely, attacks generated by G07 (AudioLDM2 configured as an ATA model) prove consistently easier to detect across most submissions. Even mid-ranking systems maintain EERs below 4.00\% on this generator. The relative ease of detecting ATA-generated deepfakes may stem from the residual acoustic artifacts introduced when conditioning the generation process directly on acoustic features, rather than textual prompts. However, the performance gap between the best primary system (S01: 0.20\%) and the baseline (B01: 5.20\%) remains significant, underscoring that while the attack is weaker, standard models are still susceptible.

For Track 2, the black-box VTA generators (G08 FoleyCrafter and G09 DiffFoley) pose an entirely different challenge. The results in Figure~\ref{fig:eers} show that while top systems like M01 (DFKI) and M02 (AHU) achieve excellent overall performance, there is noticeable variance between the two VTA models. For example, B01 exhibits an EER of 17.01\% on G08 but only 2.97\% on G09. This inconsistency emphasizes the unpredictability of black-box evaluation. The superior performance of M01 (0.35\% on G08 and 0.15\% on G09) can be attributed to its use of the BiCrossMamba architecture and large-scale external real data (AudioCaps), which likely prevents the model from overfitting to the limited 1\% black-box training data. Overall, these findings reveal that while modern generative models present severe threats, detection systems equipped with diverse pre-trained features and ensemble strategies can achieve highly reliable generalization.

\section{Insights and Future Directions}
The ESDD challenge marks a significant milestone in expanding audio anti-spoofing beyond traditional human voice. Here, we outline several key insights and future research directions.

\textbf{Component-Level Detection:}
Current ESDD systems mainly perform holistic clip-level classification. However, real-world audio is typically a mixture of various sound events~\cite{zhang2026compspoof}. As generative models advance, attackers can manipulate these components independently (e.g., altering background while preserving real speech). This motivates a shift toward component-level detection, where the model can detect the authenticity of individual acoustic sources. The recently launched ESDD2 challenge\footnote{\href{https://sites.google.com/view/esdd-challenge/esdd-challenges/esdd-2/description}{The ESDD 2 challenge page: https://sites.google.com/view/esdd-challenge/esdd-challenges/esdd-2/description}} is designed to explore this direction~\cite{zhang2026esdd2}.

\textbf{Deepfake Detection Across Audio Domains:}
While the challenge focuses solely on environmental sound, the broader audio deepfake threat spans multiple domains, such as speech, singing voice and music~\cite{huang2026audiomosaic,zaman2026deepfake,xie2026detect}. Relying on domain-specific detectors may result in fragmented solutions with limited cross-domain generalization. An important direction is the development of a unified, domain-agnostic deepfake detection framework capable of handling diverse audio modalities.

\textbf{Multimodal Audio-Visual Deepfake Detection:}
The introduction of VTA generation in Track 2 highlights a critical problem: soundscapes are often synthesized to match manipulated video content. Future detection systems can investigate the synchronization, semantic alignment, and cross-modal consistency between audio and visual streams.

\section{Conclusions}
This paper provides a comprehensive analysis of the first ESDD challenge, covering task design, datasets, baselines, and detailed evaluation across unseen and black-box generators. The results reveal that while high-fidelity generative models can significantly degrade conventional baselines (particularly under unseen generator conditions), robust generalization can still be achievable through large-scale self-supervised representations, carefully designed augmentation strategies, and ensemble modeling. The challenge highlights both the progress and remaining gaps in ESDD, and establishes a benchmark foundation for future research toward more generalizable and real-world robust audio deepfake detection systems.

\section{Generative AI Use Disclosure}
We use generative AI tools for polishing the manuscript, e.g., correcting the grammar.

\bibliographystyle{IEEEtran}
\bibliography{mybib}

\end{document}